\documentclass[%
 aip,
 amsmath,amssymb,
 reprint,%
]{revtex4-1}

\usepackage{graphicx}
\usepackage{dcolumn}
\usepackage{bm}

\usepackage[utf8]{inputenc}
\usepackage[T1]{fontenc}
\usepackage{mathptmx}

\newcommand{\sgn}{\mathop{\rm sgn}\nolimits}

\def\be{\begin{equation}}
\def\ee{\end{equation}}

\begin{document}

\preprint{AIP/123-QED}

\title[Mean-Field Failure in Interacting SCQD]{Failure of the mean-field description of magnetic fluctuations in the superconducting quantum dot}

\author{V\'aclav  Jani\v{s} }
\email{janis@fzu.cz} \author{Jiawei Yan} 

\affiliation{Institute of Physics, The Czech Academy of Sciences, Na Slovance 2, CZ-18221 Praha  8,  Czech Republic}

\date{\today}

\begin{abstract}
The zero-temperature physics of interacting quantum dots attached to superconducting leads is now well understood. The overall qualitative picture is obtained from the static mean-field approximation. The situation drastically changes at non-zero temperatures. No reliable solutions apart from numerical simulations exist there. We show that any static mean-field approximation fails at non-zero temperatures since magnetic fluctuations induce dynamical corrections that lead to broadening of the in-gap state energies to energy bands.  Spin-symmetric equilibrium state at non-zero temperatures is unstable with respect to magnetic fluctuations and the zero magnetic field can be reached only as a weak limit of the spin-polarized solution like in a magnetically ordered phase.    
\end{abstract}

\maketitle 
\section{Introduction}

Nanostructures attached to leads with specific properties display interesting and important quantum effects at low temperatures. When the impurity atoms with unpaired correlated electrons are placed in metals one observes the Kondo effect.\cite{Goldhaber:1998aa,Goldhaber-Gordon:1998aa,Cronenwett:1998aa} The correlated quantum nanostructres attached to superconductors represent  tunable microscopic Josephson junctions.\cite{Kasumov99,Kasumov03,DeFranceschi10} The simultaneous presence of strong electron correlations on semiconducting impurities and proximity of superconductors allow us to observe and analyze the interplay between the Kondo effect and the formation of the Cooper pairs carrying the Josephson current through the semiconducting nanodevices.\cite{Matsuura77,Glazman89,Rozhkov:2000aa,Buitelaar02,Aono:2004aa,Graber04,Siano04,Choi04,vanDam06,Cleuziou06,Jorgensen07,Tanaka07,Grove07,Lim08,Karrasch08,Eichler09,Yamada:2010aa,Yamada:2011aa,Luitz12,Oguri13}  

The superconducting leads induce a superconducting gap on the impurity. We then have to distinguish two types of states of the quantum dot attached to superconductors. The band states of the superconducting leads and isolated in-gap states of the dot. The band states are responsible for the Kondo effect in the strong-coupling limit at half filling, while the in-gap states play a fundamental role at the zero-temperature $0-\pi$ transition from a spin singlet to a spin doublet. Both effects have features of local quantum criticality.    

The simplest way to qualitatively simulate the critical behavior and phase transitions is to use a mean-field approximation. It cannot capture the Kondo effect caused by dynamical quantum fluctuations but it gives reasonably good quantitative predictions for weak and moderate coupling.\cite{Vecino:2003aa,Martin:2012aa}  It is, however, conceptually unacceptable at non-zero temperatures, since it binds the $0-\pi$ transition with a spurious transition to the magnetic state.\cite{Meden:2019aa}  

Dynamical corrections to the static mean-field calculated from the perturbation expansion in the interaction strength improve the results and are close to those from  the numerical renormalization group but it is applicable only in the weak-coupling $0$-phase and at zero temperature.\cite{Zonda:2016aa} One needs a consistent mean-field approximation in strong coupling ($\pi$-phase) and at non-zero temperatures to calculate dynamical effects in the superconducting quantum dot.   

The flaw of the weak-coupling mean-field approximation leading to an unphysical local phase transition can be removed by an improved mean-field approach with a two-particle self-consistency.\cite{Janis:2007aa,Janis:2008ab,Janis:2017aa,Janis:2017ab,Janis:2019aa} This mean-field approximation reproduces the Kondo strong-coupling regime when the dot is attached to metallic leads and qualitatively correctly reproduces the phase diagram of the superconducting quantum dot at all temperatures.\cite{Janis:2021aa} It hence offers a suitable starting point for including dynamical fluctuations at non-zero temperatures.

It is the aim of this paper to show how to include dynamical fluctuations  to the mean-field approximation extended to strong coupling in Ref.~\onlinecite{Janis:2021aa} at non-zero temperatures. The perturbation expansion with mean-field  Green functions breaks down. Small magnetic field induces poles in the electron-hole correlation function that incur poles in the self-energy at non-zero temperatures, which is unphysical. The  Green functions  in the perturbation expansion must then be fully dynamically renormalized. The dynamical renormalization leads at non-zero temperatures to broadening of the in-gap states to bands, whereby magnetic fluctuations have the dominant contribution. The stable  spin-symmetric solution can then be reached only as a weak limit  in which the magnetic field is set zero only after all spin-polarized calculations were concluded. The temperature-induced broadening of the in-gap energies can be observed also when the quantum dot is weakly attached to a metallic lead as recently experimentally verified. \cite{Saldana:2020aa} Such a situation can be theoretically simulated by a small imaginary part added to the frequency variable in Green functions.


\section{Mean-filed solution -- static renormalizations}

\subsection{Hamiltonian of superconducting quantum dot}

The quantum dot attached to leads is standardly modelled by a Hamiltonian consisting of three parts,  
$
\mathcal{H}=\mathcal{H}_{dot}+\sum_{s=R,L}(\mathcal{H}^s_{lead}+\mathcal{H}^s_c)
$.  
The dot Hamiltonian $\mathcal{H}_{dot}$ is approximated by an atom with a single energy level $\pm\epsilon$ for an electron (a hole) and Coulomb repulsion $U$ that in the Zeeman magnetic field $h$ reads
\begin{subequations}
\begin{equation}
\mathcal{H}_{dot}=\sum_{\sigma=\pm 1}\left(\epsilon - \sigma h\right)  d_\sigma^\dagger d_\sigma^{\phantom{\dag}}
+Ud_\uparrow^\dag d_\uparrow^{\phantom{\dag}} d_\downarrow^\dag d_\downarrow^{\phantom{\dag}} \,.
\end{equation}
Here $\sigma =\pm 1$ corresponds to spin up/down.
The Hamiltonians of the leads
\begin{equation}
\mathcal{H}^s_{lead}=\sum_{\mathbf{k}\sigma}
\epsilon(\mathbf{k})c_{s\mathbf{k}\sigma}^\dag c_{s\mathbf{k}\sigma}^{\phantom{\dag}} 
-\Delta_{s}\sum_\mathbf{k}(e^{i\Phi_s}
c_{s\mathbf{k}\uparrow}^\dag c_{s\mathbf{-k}\downarrow}^\dag+\textrm{H.c.})\end{equation}
\end{subequations}
represent BCS superconductors with gaps $\Delta_{s}$ and $s = L,R$ denoting left, right lead. Finally, the hybridization term for the contacts reads 
$
\mathcal{H}^s_c=-t_s\sum_{\mathbf{k}\sigma}
(c_{s\mathbf{k}\sigma}^\dag d_\sigma^{\phantom{\dag}}+\textrm{H.c.}) 
$.
We will use identical left and right hybridizations and superconductors, $\Delta_{L} = \Delta_{R}=\Delta$ and $t_{L} = t_{R} = t$  without loss of generality. The asymmetric situation can be transformed to a symmetric one.\cite{Kadlecova:2017aa}
We will further approximate the Green function in the leads by its value  at the Fermi energy and denote $\Gamma= 2\pi t^{2}\rho_{0}$ being the effective hybridization strength. We denoted $\Phi = \Phi_{L} - \Phi_{R}$ the difference between the phases of the attached superconducting leads and $\rho_{0}$ the density of states of the lead electrons at the Fermi energy.  

\subsection{Mean-field one-electron propagators}

The noninteracting quantum dot is exactly solvable. Its solution introduces band states with a continuous spectrum and isolated sharp in-gap states. A mean-field solution of the interacting dot leads to the same spectral structure with only statically renormalized parameters. The general form of the normal mean-field propagator on the dot is 
\begin{subequations}
\begin{align}
G_{\sigma}(\omega_{+}) &= \frac{\omega\left[1 + s(\omega_{+})\right] + \sigma \bar{h} + \bar{\epsilon}}{D_{\sigma}(\omega_{+})} \,.
\end{align}
The anomalous propagator then is
\begin{align}
\mathcal{G}_{\sigma}(\omega_{+}) &= - \frac{c_{\Phi}\left[s(\omega_{+}) \Delta - U\nu\right] }{D_{\sigma}(\omega_{+})} \,, 
\end{align}
\end{subequations}
where $c_{\Phi}= \cos\left(\Phi/2\right)$. We abbreviated $\omega_{+} = \omega + i0^{+}$ to indicate the way the real axis is reached from the complex plane.
The denominator in the mean-field approximation is a quadratic form
\begin{multline}\label{eq:D-SpinPolarized}
D_{\sigma}(\omega)= \left[\omega_{+}\left(1 + s(\omega_{+})\right) + \sigma\bar{h}\right]^{2}
\\
 -\ \bar{\epsilon}^{2} - c_{\Phi}^{2}\left[s(\omega_{+}) \Delta - U\nu\right]^{2} \,.
\end{multline}
The impact of the lead electrons on the dot propagator is embodied in a hybridization self-energy with the following analytic properties
\begin{subequations}
\begin{align}
s(x\pm i0) &=\pm\frac{i\Gamma\sgn(x)}{\sqrt{x^2-\Delta^2}}\qquad\mathrm{ for }\qquad|x|>\Delta\ ,
\\
s(x\pm i0) &=\phantom{\pm}\frac{\Gamma}{\sqrt{\Delta^2-x^2}}\qquad\mathrm{ for }\qquad|x|<\Delta\ .
\end{align}
\end{subequations}
We further introduced static renormalizations of the impurity energy  
$
\bar{\epsilon} = \epsilon +  U n/2
$ and of the magnetic field 
$
\bar{h} = h + \Lambda m/2$. 
The  renormalization parameters $n,m$ stand for the dot  charge and spin and  densities, respectively. They are determined from the corresponding mean-field equations with the normal Green functions $G_{\sigma}$. The Cooper-pair density $\nu$ is proportional to the spin-symmetric sum of the anomalous Green functions $\mathcal{G}_{\sigma}$. Notice that we introduced a renormalized static effective interaction $\Lambda$ controlling the response to the magnetic field. It replaces the bare interaction so that to suppress the impurity magnetic transition of the weak-coupling mean-field approximation. It will be connected with the two-particle vertex in the following subsection. 

The full propagator can be split into two parts, the gap, $\omega\in (-\Delta,\Delta)$, and band, $|\omega|>\Delta$, contributions, $G_{\sigma}(\omega) = G_{\sigma}^{g}(\omega) + G_{\sigma}^{b}(\omega)$. The separation is made according to their non-zero imaginary parts. The corresponding real parts are calculated from the Kramers-Kronig relation that extend on the whole frequency interval. In what follows we will concentrate on the propagator due to the isolated in-gap states of the mean-field solution.  Its normal and anomalous parts in the mean-field approximation are 
\begin{subequations}
\begin{align}
G_{\sigma}^{g}(\omega) & = \frac{X_{\sigma} + \bar{\epsilon}}{2K_{\sigma}(\omega - \omega_{\sigma})} + \frac{X_{-\sigma} - \bar{\epsilon}}{2K_{-\sigma}(\omega + \omega_{-\sigma})}\,,
\\
\mathcal{G}_{\sigma}^{g}(\omega) & = - \frac{c_{\Phi}}2\left[\frac{s_{\sigma}\Delta - U\nu}{K_{\sigma}(\omega - \omega_{\sigma})} - \frac{s_{-\sigma}\Delta - U\nu}{K_{-\sigma}(\omega + \omega_{-\sigma})}\right]\,.
\end{align}
\end{subequations}
We denoted the zeros of the denominator, the poles of the gap propagator  
\begin{equation}\label{eq:omega-h}
\omega_{\sigma}(1+ s_{\sigma}) = - \sigma \bar{h} 
+ \sqrt{\bar{\epsilon}^{2} + c_{\Phi}^{2}\left(s_{\sigma}\Delta - U\nu\right)^{2}} \,,
\end{equation}
with $s_{\sigma} = s(\omega_{\sigma})$, used the renormalized pair energy
\begin{subequations}
 \begin{align}
X_{\sigma} &= \sqrt{\bar{\epsilon}^{2} + c_{\Phi}^{2} \left(s_{\sigma}\Delta  - U\nu\right)^{2}}  \,,
\end{align}
and the residue of the poles
\begin{multline}
K_{\sigma} =  X_{\sigma}\left[1 + \frac{\Delta^{2}s_{\sigma}}{\Delta^{2} - \omega_{\sigma}^{2}} \right] - c_{\Phi}^{2}\left(s_{\sigma}\Delta - U\nu\right)\frac{\omega_{\sigma} s_{\sigma}\Delta}{\Delta^{2} - \omega_{\sigma}^{2}} \,.
 \end{multline}
 \end{subequations} 

\subsection{Mean-field two-particle vertex}

The standard weak-coupling mean-field theory introduces does not affect the interaction strength. It is not enough in intermediate and strong coupling where magnetic fluctuations may drive the system to a magnetic state. One has to introduce a renormalization of the interaction in the magnetic response to tame the dybnamical fluctuations. It is achieved by simplified parquet equations with a two-particle self-consistency.\cite{Janis:2017aa,Janis:2017ab} The resulting effective interaction, approximate static electron-hole irreducible vertex,  replacing the bare interaction in the magnetic response obeys the following self-consistent equation\cite{Janis:2019aa,Janis:2021aa}  
\begin{equation}\label{eq:Lambda-general}
\Lambda =   \frac {U n_{\uparrow}n_{\downarrow}  }{ n_{\uparrow}n_{\downarrow}  + \Lambda^{2} \mathcal{X}} 
\end{equation}
where $n_{\sigma}$ is the density electrons with spin $\sigma$ and 
\begin{equation}\label{eq:X-integral}
\mathcal{X} =  -\frac {1}{\beta} \sum_{\nu_{m}}\frac{ \psi(i\nu_{m}) \psi(-i\nu_{m})\phi(-i\nu_{m})}{1 + \Lambda\phi(-i\nu_{m})} 
\end{equation}
is a screening integral hindering the spurious transition to the magnetic state. 
We introduced a spin-symmetric electron-hole bubble 
\begin{subequations}
\begin{multline}\label{eq:eh-bubble}
\phi(i\nu_{m}) = \frac 1{2\beta}\sum_{\sigma}\sum_{\omega_{n}}\left[G_{\bar{\sigma}}(i\omega_{n} + i\nu_{m}) G_{\sigma}(i\omega_{n}) 
\right. \\ \left.
+\ \mathcal{G}_{\bar{\sigma}}(i\omega_{n} + i\nu_{m}) \mathcal{G}_{\sigma}(i\omega_{n}) \right] 
\end{multline}  
and the electron-electron bubble
\begin{equation}
\psi(i\nu_{m}) =  \frac 1{\beta } \sum_{\omega_{n}} G_{\downarrow}(i\nu_{m} - i\omega_{n})G_{\uparrow}(i\omega_{n})  \,.
\end{equation}
\end{subequations}
The electron-electron bubble does not contain anomalous propagators. We denoted $\omega_{n} = 2(n + 1)\pi/\beta$ and $\nu_{m}=2m \pi/\beta$ the fermionic and bosonic Matsubara frequencies, respectively. We do not need to use analytic continuation to real frequencies to determine the effective interaction. Unlike the self-energy where we must separately analyze the contributions from  the in-gap and band states.

\section{Dynamical fluctuations - breakdown of the mean-field approximation}

The two-particle scatterings contain the dominant dynamical corrections to the static mean-field solution. The electron-hole scatterings drive the system towards a transition to the magnetic state while the electron-electron scatterings slow down this process by screening the bare interaction. Both generic two-particle bubbles at zero temperature are gapped and free of poles in the spin-symmetric state. A small magnetic field induces poles in the electron-hole bubble, which makes the spin-symmetric solution unstable with respect to magnetic fluctuations  everywhere at non-zero temperatures and in the $\pi$-phase at zero temperature.     

\subsection{Magnetic fluctuations}

The instability of the spin-symmetric state with respect to magnetic fluctuations is caused by sensitivity of the in-gap states to the Zeeman field. The imaginary part of the electron-hole bubble from Eq.~\eqref{eq:eh-bubble}  of the spin-polarized solution in the mean-field approximation of the preceding section analytically continued to real frequencies has the following magnetically induced singular part
\begin{multline}\label{eq:Im-Phi-g}
\Im\phi_{g}(\omega_{+}) =  \frac {\pi\left(\Delta f_{\downarrow} - \Delta f_{\uparrow}\right)}{4K_{\uparrow}K_{\downarrow}}\left[X_{\downarrow}X_{\uparrow} + \bar{\epsilon}^{2} + c_{\Phi}^{2}\left(s_{\uparrow}\Delta - U\nu\right) 
\right. \\ \left.
\times\left(s_{\downarrow}\Delta - U\nu\right)\right]\left[\delta(\omega + \Delta\omega) - \delta(\omega - \Delta\omega)\right] \,. 
\end{multline}
We used the Fermi function $f(x) = 1/(e^{\beta x} + 1)$ and denoted $\Delta f_{\sigma} = f(-\omega_{\sigma}) - f(\omega_{\sigma}) = \tanh(\beta \omega_{\sigma}/2)$, $\Delta\omega = \omega_{\downarrow} - \omega_{\uparrow}$.  Although the poles in the electron-hole bubble vanish at zero magnetic field, they make the static spin-symmetric self-energy unstable at non-zero temperatures.


\subsection{Dynamical self-energy}

The dynamical correction to the static self-energy is determined from the Schwinger-Dyson equation. The dynamical (spectral) self-energy must share the same critical behavior with the mean-field one. That is, only the spin-symmetric part of the self-energy enters the Schwinger-Dyson equation.\cite{Janis:2021aa}  We split the spin-symmetric spectral self-energy at non-zero temperatures into two contributions, $\Sigma^{Sp}(\omega_{+}) = \Sigma^{Sp}_{0}(\omega_{+}) + \Sigma^{Sp}_{T}(\omega_{+})$, and analogously for the anomalous part. The first one  
\begin{subequations}\label{eq:SD-Sigma0}
\begin{multline}
\Sigma^{Sp}_{0}(\omega_{+}) = U\Lambda \int_{-\infty}^{\infty} \frac{d x}{\pi} f(x)\left\{ \frac{\phi(x - \omega_{+})}{1 + \Lambda \phi(x - \omega_{+})}  
\right. \\ \left.
\times \Im \bar{G}^{Sp}(x_{+}) +\bar{G}^{Sp}(\omega_{+} + x) \Im\left[\frac{\phi(x_{+})}{1 + \Lambda\phi(x_{+})} \right]\right\} \,.
\end{multline}
and analogously the anomalous self-energy 
\begin{multline}
c_{\Phi}\mathcal{S}^{Sp}_{0}(\omega_{+}) = U\Lambda \int_{-\infty}^{\infty} \frac{d x}{\pi} f(x) \left\{\frac{\phi(x - \omega_{+})}{1 + \Lambda \phi(x - \omega_{+})}  
\right. \\ \left.
\times \Im \bar{\mathcal{G}}^{Sp}(x_{+}) + \bar{\mathcal{G}}^{Sp}(\omega_{+} + x) \Im\left[\frac{\phi(x_{+})}{1 + \Lambda\phi(x_{+})} \right]\right\} 
\end{multline}
\end{subequations}
survive to zero temperature. The second part is temperature-induced and reads
%
\begin{multline}\label{eq:SD-SigmaT}
\Sigma^{Sp}_{T}(\omega_{+}) = - TU\Lambda \int_{-\infty}^{\infty} \frac{d x}{\pi}\frac{\bar{G}^{Sp}(\omega_{+} + Tx)}{\sinh(x)} 
\\
\times\Im\left[\frac{\phi(Tx_{+})}{1 + \Lambda\phi(Tx_{+})} \right] \,.
\end{multline}
The same holds for the anomalous self-energy $c_{\Phi}\mathcal{S}^{Sp}_{T}(\omega_{+}) $ with the normal propagator replaced by the anomalous one. 
We rescaled the integration variables to better follow the low-temperature dependence of the temperature-induced self-energy.

The one-particle propagator on the right-hand side is spin symmetric, hence $\bar{G}^{Sp}(x_{+}) =\left(G^{Sp}_{\uparrow}(x_{+}) + G^{Sp}_{\downarrow}(x_{+})\right)/2$ and $\bar{\mathcal{G}}^{Sp}(x_{+}) =\left(\mathcal{G}^{Sp}_{\uparrow}(x_{+}) + \mathcal{G}^{Sp}_{\downarrow}(x_{+})\right)/2$. 
The explicit form of $G^{Sp}_{\sigma}$ and $\mathcal{G}^{Sp}_{\sigma}$ depends on the  level of self-consistency used in the Schwinger-Dyson equation. They are  $G_{\sigma}$ and $\mathcal{G}_{\sigma}$ from the preceding section in the mean-field approximation. Such a choice would, however, lead to poles in the self-energy in the spin-polarized solution, which is unphysical. One must hence introduce a full dynamical one-particle self-consistency in the Schwinger-Dyson equation to suppress the mean-field singularities in the spectral self-energy. It means that $G^{SP}$ and $\mathcal{G}^{Sp}$ on the right-hand side of Eqs.~\eqref{eq:SD-Sigma0} and~\eqref{eq:SD-SigmaT} contain the spectral self-energies $\Sigma^{Sp}$ and $\mathcal{S}^{Sp}$. The one-particle self-consistency then smoothes the poles to resonances and  leads to broadening of the in-gap-state energies to in-gap bands at non-zero temperatures. It appears that when going into strong coupling  or to high temperatures one needs to dynamically renormalize also the propagators in the two-particle bubbles of the mean-field approximation.

\section{Broadening of the in-gap states at non-zero temperatures}
 
The dominant contribution to the temperature-induced self-energies $\Sigma^{Sp}_{T}(\omega_{+})$ and $\mathcal{S}^{Sp}_{T}(\omega_{+}) $ comes from small frequencies of the integrand with the inverse of the hyperbolic sine. We can then determine this self-energy at low temperatures from simplified algebraic equations
\begin{subequations}\label{eq:SE-LT}
\begin{align}
\Sigma^{Sp}_{T}(\omega_{+}) &=  L(T) \bar{G}^{Sp}(\omega_{+}) \,,
\\
c_{\Phi}\mathcal{S}^{Sp}_{T}(\omega_{+}) &=  L(T)\bar{\mathcal{G}}^{Sp}(\omega_{+})  \,,
\end{align}
\end{subequations} 
where we introduced a thermodynamic factor 
\begin{align}\label{eq:LT-integral}
L(T) & = -TU\Lambda\int_{-\infty}^{\infty} \frac{d x}{\pi\sinh(x)} 
\Im\left[\frac{\phi(Tx_{+})}{1 + \Lambda\phi(Tx_{+})} \right] \,.
\end{align}
 We stress that the thermodynamic factor becomes non-zero only when the magnetically-induced poles in the electron-hole bubble $\phi(\omega)$ are taken into account. 
  
The thermodynamic factor $L(T)$, when the mean-field electron-hole bubble from Eq.~\eqref{eq:Im-Phi-g} is used in Eq.~\eqref{eq:LT-integral},  reduces in weak coupling, where we can resort to second-order self-energy, and in the weak limit to zero magnetic field where $\Delta \omega \to 0$ results in
\be\label{eq:LT2}
L_{2}(T) = U\Lambda f(\omega_{0})\left(1 - f(\omega_{0})\right)\frac{X(\omega_{0})^{2}}{K(\omega_{0})^{2}}\,,
\ee  
where $\omega_{0}>0$ is the pole in the mean-field propagator of the spin-symmetric state.    

We resolve the Schwinger-Dyson equations~\eqref{eq:SD-Sigma0} with the dynamically renormalized one-electron propagators.  The spin-symmetric version of the denominator $D_{\sigma}(\omega_{+})$ of the on-particle Green functions  from Eq.~\eqref{eq:D-SpinPolarized} at half filling, $\bar{\epsilon} = 0$, is
\begin{multline}\label{eq:D-SpinSymmetric}
D^{Sp}(\omega_{+}) = \left[\bar{\omega}_{+} - \Sigma(\omega_{+})\right]^{2}
- c_{\Phi}^{2}\left(\bar{s}(\omega_{+}) - \mathcal{S}(\omega_{+})\right)^{2}   \,.
\end{multline}
It contains  the full normal and anomalous self-energies $\Sigma(\omega) = \Sigma^{Sp}_{0}(\omega) + \Sigma^{Sp}_{T}(\omega)$ and $\mathcal{S}(\omega) = \mathcal{S}^{Sp}_{0}(\omega) + \mathcal{S}^{Sp}_{T}(\omega)$. We abbreviated  $\bar{\omega}_{+} = \omega_{+}\left(1 + s(\omega_{+})\right)$ and  $\bar{s}(\omega_{+}) = s(\omega_{+})\Delta - U\nu$. We can neglect $\Im\Sigma^{Sp}_{0}(\omega_{+})$ and $\Im\mathcal{S}^{Sp}_{0}(\omega_{+})$ at low-temperatures within the gap. The positions $\pm \omega_{0}$  of the in-gap states in the dynamically renormalized Green function are determined from  $\Re D^{Sp}(\omega_{0})=0$, being 
%
\begin{multline}\label{eq:L-UB}
\left[\omega_{0}\left(1 + s(\omega_{0})\right) - \Re\Sigma_{0}(\omega_{0})\right]^{2}
\\
 = c_{\Phi}^{2}\left[s(\omega_{0})\Delta  - U\nu - \Re\mathcal{S}_{0}(\omega_{0})\right]^{2} \,.
\end{multline} 
The zero-temperature poles  at $\pm \omega_{0}$ are broadened at non-zero temperatures to bands  with $\Im D^{Sp}(\omega_{+})\neq 0$. Solving Eq.~\eqref{eq:SD-SigmaT} in the low-temperature limit with $\Im\Sigma_{0}(\omega_{+})=\Im\mathcal{S}_{0}(\omega_{+})=0$ and with the asymptotic form \eqref{eq:SE-LT} we obtain for the centers of the in-gap bands,  maximum of $\Im D^{Sp}(\omega)$, 
\begin{multline}
\Im D^{Sp}(\omega_{0})^{2} =L(T) \left[\left( 2 c_{\Phi}^{2}\bar{s}(\omega_{0}) - \Re\mathcal{S}_{0}(\omega_{0})\right)^{2} - L(T)
\right. \\ \left. 
+\ \sqrt{1 - \frac{L(T)^{2}}{\left( 2 c_{\Phi}^{2}\left(\bar{s}(\omega_{0}) - \Re\mathcal{S}_{0}(\omega_{0})\right)^{2} - L(T)\right)^{2}}} \right] \,,
\end{multline}
which is positive at non-zero temperatures, meaning there are no zeros in the denominator, no poles, of the Green function  for $T>0$. 

\begin{figure}
\includegraphics[width=7cm]{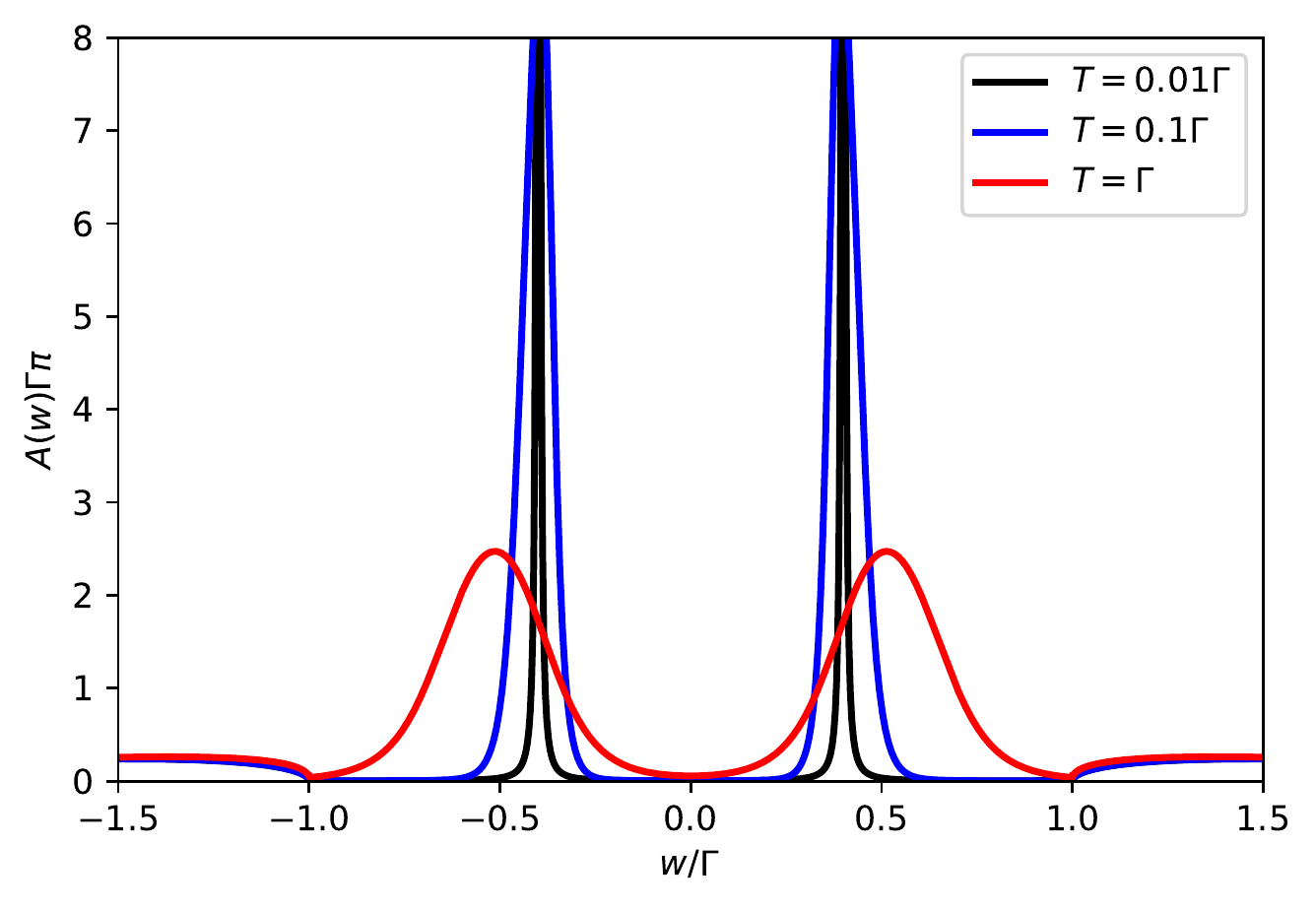}
\caption{Normal part of the spectral function of the superconducting quantum dot  in weak coupling, $U=\Gamma$, at half filling,  with the superconducting gap $\Delta=\Gamma$, and for the phase difference $\Phi=0$ calculated for increasing temperatures with second-order spectral self-energy from Eqs.~\eqref{eq:SD-Sigma0} and Eq.~\eqref{eq:SD-SigmaT} calculated with the fully renormalized Green functions. The Cooper-pair singlet density $\nu$ decreases with temperature as more weight from the doublet $\pi$ phase is getting admixed, $\nu=0.21,0.20,0.03,$ for $T=0.01\Gamma,0.1\Gamma,\Gamma$, respectively. \label{fig:fullSigma} }
\end{figure}  
 
\begin{figure}
\includegraphics[width=7cm]{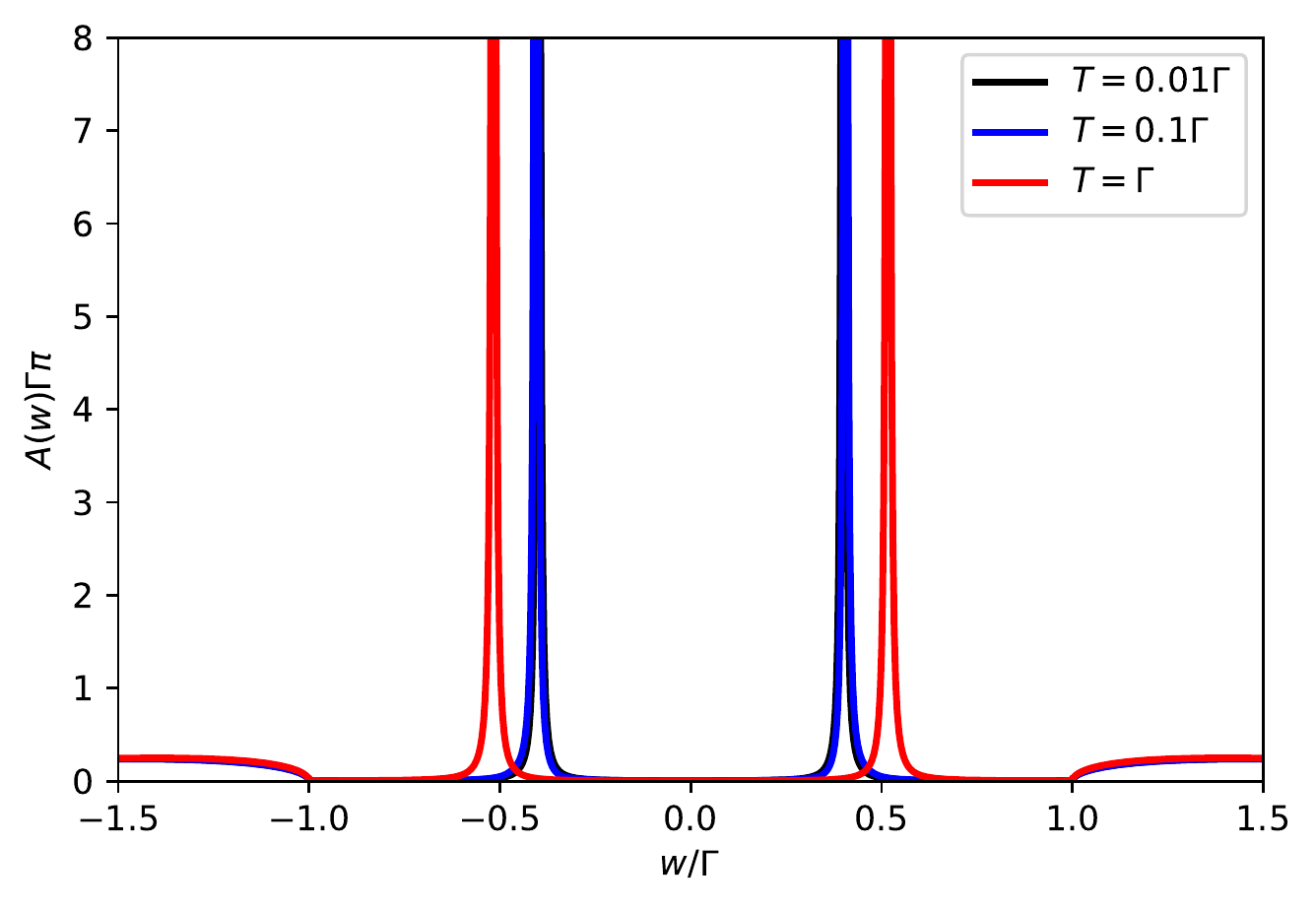}
\caption{Normal part of the spectral function of the superconducting quantum dot  in the setting of Fig.~\ref{fig:fullSigma} calculated for increasing temperatures with second-order spectral self-energy only from Eqs.~\eqref{eq:SD-Sigma0}. There is negligible broadening at low temperatures due to $\Im\Sigma^{Sp}_{0}(\omega_{+})$ compared $\Im\Sigma^{Sp}_{T}(\omega_{+})$.  \label{fig:Sigma0} }
\end{figure}

 \begin{figure}
\includegraphics[width=7cm]{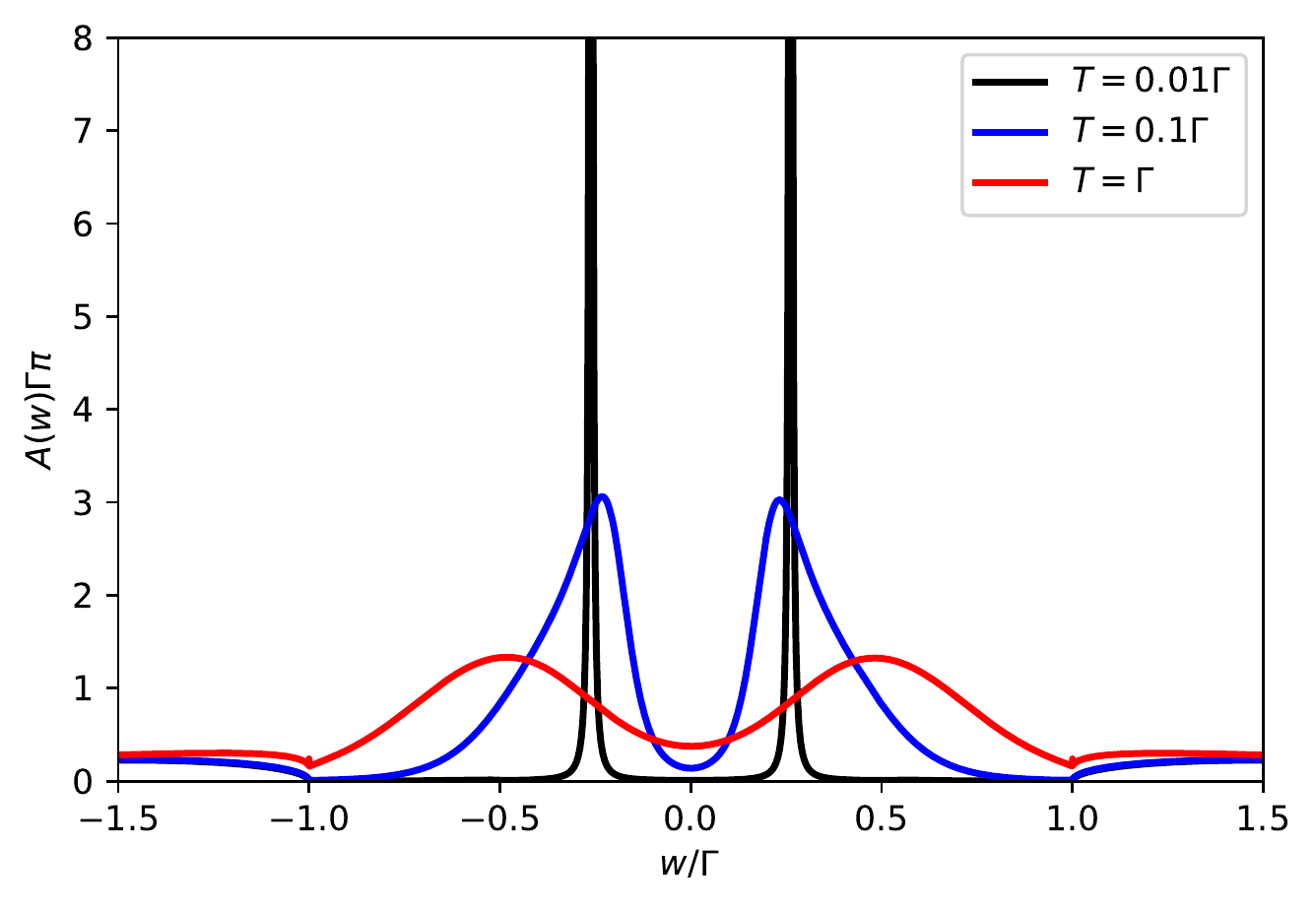}
\caption{Normal part of the spectral function of the superconducting quantum dot  in intermediate coupling, $U= 2\Gamma$, at half filling, with the superconducting gap $\Delta=\Gamma$, and for the phase difference $\Phi=0$ as in Fig~\ref{fig:fullSigma}. The broadening gets stronger and the centers of the in-gap bands are driven faster away from the Fermi energy with increasing temperature.    \label{fig:Sigma2} }
\end{figure}  
The precise form of the in-gap bands can be determined only numerically. The temperature-dependent spectral function for $U=\Gamma$ is plotted in Figs.~\ref{fig:fullSigma} and~\ref{fig:Sigma0}. The first plot includes the contributions from both self-energies $\Sigma^{Sp}_{0}(\omega)$ and $\Sigma^{Sp}_{T}(\omega)$, while the second only from $\Sigma^{Sp}_{0}(\omega)$.  We used the dynamically renormalized both one and two-particle Green functions in the Schwinger-Dyson equations~\eqref{eq:SD-Sigma0} and~\eqref{eq:SD-SigmaT} with a small regularizing imaginary part added to the frequency. We used second-order spectral self-energy with the effective interaction only weakly screened,  $\Lambda \in( 0.98\Gamma, 0.99 \Gamma)$ in the plotted temperature range. The thermodynamic factor $L_{2}(T)$ from Eq.~\eqref{eq:LT2} was taken as a starting value in the iterations to include magnetic fluctuations in $\Sigma^{Sp}_{T}(\omega)$. Comparing the two figures we can clearly see that the magnetically incited broadening  of the zero-temperature poles of the normal Green function due to $\Sigma^{Sp}_{T}(\omega)$ dominates and grows rapidly with increasing temperature. We also plotted in Fig.~\ref{fig:Sigma2} the broadening of the in-gap energies in intermediate coupling for $U=2\Gamma$ where the in-gap bands overlap. The renormalization of the propagators in the electron-hole bubble is then mandatory  to suppress the spurious increase of the density of states at the Fermi energy of the bare (mean-field) bubble.  The unrenormalized bubble does not guarantee charge conservation, which causes the unphysical behavior of the spectral function around the Fermi energy.  The anomalous part of the spectral function has a similar form with only negative values for positive frequencies. The dynamical fluctuations and temperature drive the centers of the in-gap bands towards the band edges.   


\section{Conclusions}

The mean-field spin-symmetric solution becomes unstable with respect to magnetic fluctuations. The magnetic fluctuations cannot be neglected, which means that zero magnetic field must be reached  as a weak limit of non-zero values. A stable spin-polarized solution can, however, be obtained only if the Green functions in determining the self-energy are dynamically renormalized. Their dynamical renormalization leads to smearing of the poles of the zero-temperature Green function. The isolated in-gap states get broadened with increasing temperature. Consequently, any approximation with static renormalizations (mean-field) fails at non-zero temperatures and the dynamical spectral self-energy  from the Schwinger-Dyson equation with the fully renormalized one-particle propagator must be used to stabilize the equilibrium state. Otherwise, unphysical behavior of the spectral self-energy with poles at non-zero temperatures is obtained. This is a generic feature when isolated in-gap states coexist with continuous band states at zero temperature.

\section*{Acknowledgment}
 The research was supported by Grant No. 19-13525S of the Czech Science Foundation and INTER-COST LTC19045 Program and the COST Action NANOCOHYBRI (Grant No. CA16218) of the Czech Ministry of Education, Youth and Sports. 
 
 \section*{Data Availability}
The data that supports the findings of this study are available within the article. 
%

%

\end{document}